\documentclass[10pt,letterpaper,typeset,superscriptaddress,twocolumn]{revtex4}

\usepackage{amsfonts}
\usepackage{amsmath}
\usepackage{amssymb}
\usepackage{graphicx}

\begin{document}

\title{Quantum tomography via equidistant states}
\author{C. Paiva-S\'anchez}
\author{E. Burgos-Inostroza}
\author{O. Jim\'enez}
\author{A. Delgado}
\affiliation{Center for Optics and Photonics, Universidad de Concepci\'{o}n,
Casilla 4016, Concepci\'{o}n, Chile.}
\affiliation{Departamento de Fisica, Universidad de Concepci\'on, Casilla
160-C, Concepci\'on, Chile.}

\begin{abstract}

We study the possibility of performing quantum state tomography via equidistant states. This class of states allows us to propose a non-symmetric informationally complete POVM based tomographic scheme. The scheme is defined for odd dimensions and involves an inversion which can be analytically carried out by Fourier transform.

\end{abstract}

\date{\today}
\maketitle

\section{Introduction}
\label{sec1}

The determination of quantum states and processes \cite{Paris} plays an important role in the foundation of quantum theory \cite{Breitenbach,Ourjoumtsev} and quantum information theory \cite{Miquel,OBrien,Walther,Haeffner}. Quantum states are mathematical descriptions of physical systems, which provide all the information necessary for characterizing the result of any possible experiment. According to this, quantum states, unlike their classical counterparts, are not observable; therefore the determination of quantum states becomes an important problem.  Furthermore, in quantum information theory the implementation of processes requires a quality assessment, which typically bases on the determination of quantum states.

The most commonly used method to determine unknown quantum states belonging to a finite $N$-dimensional Hilbert space is quantum tomography \cite{DAriano1,Dariano2}. This method requires the preparation of a large ensemble of copies of the unknown state and the capacity of measuring transition probabilities toward a set of states, which when properly chosen leads to the determination of the unknown quantum state. In general, the coefficients of the unknown state are lineal functions of the transition probabilities and thus an inversion is required. Due to this inversion and the inherent experimental inaccuracies, the reconstructed mathematical object might violate the positivity constraint defining quantum state. To overcome this problem, techniques like maximum likelihood estimation \cite{Hradil,Tan} or Bayesian analysis are applied.

Several schemes for quantum tomography have been proposed depending on the choice of the set of states. Standard quantum tomography bases on the measurement of transition probabilities, calculated according to Born statistical formula, toward the $N$ eigenstates of a $N$-dimensional representation of the $N^2-1$ generators of $SU(N)$ \cite{James,Thew}. A second alternative is symmetric informationally complete (SIC) POVM based quantum tomography \cite{Prugovecki,Flammia,Renes,Durt}, which resorts to the measurement of transition probabilities toward a set of $N^2$ linearly dependent states such that the absolute value of the inner product between two of them is $1/\sqrt{N+1}$. The existence of SIC-POVM has been shown analytically for $N\le8$ and numerically for $N\le47$.  The existence of SIC-POVM in arbitrary, finite dimensions has been conjectured but not proven. A third alternative is quantum tomography based on mutually unbiased bases (MUB) \cite{Schwinger,Ivanovic,Wootters}, which resorts to the measurement of transition probabilities to states belonging to $N+1$ $N$-dimensional orthonormal bases. States belonging to different bases have an absolute value of the inner product equal to $1/\sqrt{N}$ \cite{Adamson,Bandyopadhyay}. It has been shown that MUBs exist when $N$ is an integer power of a prime. Otherwise the existence of MUBs is an open problem.

In this article we introduce a new scheme for implementing quantum tomography. This is based on the so-called equidistant states \cite{Roa}. These are such that the inner product between two of them is a complex constant or its conjugate. Projectors onto these states allow us to define a complete POVM which reconstructs an unknown density operator acting on a Hilbert space of finite, odd dimension $N$.  The POVM can be analytically constructed for these dimensions being the number of measurements to be performed equal to $N^2$, as few as in the case of SIC-POVM or MUB-POVM based quantum tomography.  The scheme requires inversion. However, this is reduced to the inversion of $(N+1)/2$ circulant $N\times N$ matrices, which can be performed by Fourier transform. Thereby, this new scheme lays between standard quantum tomography and POVM based quantum tomography.

\section{Equidistant States}
\label{sec2}

Let us start by reviewing the set ${\cal B}_0(\alpha)$ of equidistant states. The $N$ non-orthogonal states  $|\alpha_j\rangle$ ($j=0,\dots,N-1$) in this set belong to a $N$-dimensional Hilbert space $\mathcal H$ and are defined by the property
\begin{equation}
\langle\alpha_j|\alpha_{j^\prime}\rangle=|\alpha|e^{i\theta},\hspace{0.035in}\forall\hspace{0.05in} j>j^{\prime},
\end{equation}
that is the inner product between any two equidistant states is equal to $\alpha$ or its conjugate $\alpha^*$. The states in ${\cal B}_0(\alpha)$ are linearly independent when the modulo of $\alpha$ is constrained to the interval $[0,|\overline{\alpha}_\theta|)$, where $|\overline{\alpha}_\theta|$ is a function of $\theta$ and $N$ given by
\begin{equation}
|\overline{\alpha}_\theta|=\frac{\sin(\frac{\pi-\theta}{N})}{\sin\left(\theta+\frac{\pi-\theta}{N}\right)},
\label{absalpha2}
\end{equation}
which is the largest possible value of the inner product $|\alpha|$ for $\theta$ and $N$ fix. The states in ${\cal B}_0(\alpha)$ become linearly dependent when $|\alpha|=|\overline{\alpha}_\theta|$ for $\theta\neq 0$, lying symmetrically on a $(N-1)$-dimensional subspace. In the case $\theta=0$ the value of the bound  $|\overline{\alpha}_\theta|$ becomes one. Consequently, states with $\theta=0$ and $|\overline{\alpha}_\theta|$=1 are all within the same ray.

The set of equidistant states ${\cal B}_0(\alpha)$ becomes a set of symmetric states when the inner product $\alpha$ is real \cite{Roa}, that is $\theta=0$ or $\theta=\pi$. For these values of $\theta$ the upper bound $|\bar\alpha_\theta|$ we obtain $|\bar\alpha_0|=1$ and $|\bar\alpha_\pi|=1/(N-1)$ respectively. In general, for a complex inner product $\alpha$ equidistant states are different from symmetric states.

A canonical decomposition of equidistant states \cite{Jimenez} is the following
 \begin{equation}
 |\alpha_j\rangle=\frac{1}{\sqrt{N}}\sum_{k=0}^{N-1}\sqrt{\lambda_k}(\omega_k^{j})^*|k\rangle.
 \label{EQUIDISTANT}
 \end{equation}
The coefficients $\lambda_k$ are eigenvalues of the matrix $M$ containing the inner products $\langle\alpha_i|\alpha_j\rangle$. These are given by
\begin{equation}
 \lambda_k=1-|\alpha|\frac{\sin(\theta+\frac{k\pi-\theta}{N})}{\sin(\frac{k\pi-\theta}{N})}
 \end{equation}
 and fulfill the identity
 \begin{equation}
 \sum_{k=0}^{N-1}\lambda_k=N.
 \label{IDENTITY}
 \end{equation}
 The complex phases $\omega_k$ are the coefficients entering in the matrix $T$ which diagonalizes the matrix $M$. These phases are given by
 \begin{equation}
 \omega_k=e^{\frac{2i}{N}(\theta-k\pi)}.
 \end{equation}

 \section{Tomography}
\label{sec3}

In order to reconstruct the state of the system we consider the sets ${\cal B}_s(\alpha)$ with $s=0\dots N-1$ defined as
 \begin{equation}
{\cal B}_s(\alpha)=\{|\alpha^s_j\rangle=X^s|\alpha_j\rangle\},
\label{SETS}
 \end{equation}
 where the action of the unitary operator $X$ onto the canonical base is
 \begin{equation}
 X|k\rangle=|k\oplus1\rangle,
 \end{equation}
and the addition is performed modulo $N$. Let us note that $X^0=X^N={\bf 1}$ and that the unitarity of $X$ indicates that each set ${\cal B}_s(\alpha)$ is composed of equidistant states. Furthermore, all sets ${\cal B}_s(\alpha)$ are characterized by the same value $\alpha$ of the inner product.

 Now, defining the projectors $\Pi_j^s=|\alpha_j^s\rangle\langle\alpha_j^s|$ onto equidistant states and using Eqs. (\ref{EQUIDISTANT}),  (\ref{IDENTITY}) and (\ref{SETS}) it can be shown that
 \begin{equation}
\sum_{j,s=0}^{N-1}\Pi_j^s=N{\mathbf 1},
\label{POVM}
 \end{equation}
and thus the $N^2$ projectors $\Pi_j^s$ form a positive operator valued measurement. Thereby, the set of projectors seems to be well suited for reconstructing an unknown density matrix.

We now consider an arbitrary unknown density operator $\rho$ acting onto $\cal H$. This space is spanned via the canonical base $\{|k\rangle\}$ with $k=0\dots N-1$. Thereby, the density operator can be represented as
\begin{equation}
\rho=\sum_{k,k'=0}^{N-1} \rho_{k,k'}|k\rangle\langle k'|.
\end{equation}
The probability $P^s_j$ of projecting the density operator $\rho$ onto the state $j$-th state $|\alpha^s_j\rangle$ of the base ${\cal B}_s(\alpha)$ is calculated through the Born statistical formula
\begin{equation}
P^s_j=Tr(\rho\Pi_j^s).
\end{equation}
According to Eq. (\ref{POVM}) the set of $N^2$ probabilities $P_j^s$ obeys the constraint
\begin{equation}
\sum_{s,j=0}^{N-1} P^s_j=N.
\end{equation}
Considering the decomposition of the equidistant states onto the canonical base the probabilities $P_i^s$ become
\begin{equation}
P^s_j=\frac{1}{N}\sum_{p,q=0}^{N-1} e^{\frac{2\pi i}{N}(p-q)j}\sqrt{\lambda_{p-s}\lambda_{q-s}}\rho_{q,p},
\label{Psi0}
\end{equation}
where  the operations with subindexes are carried out modulo $N$.

The previous expression leads us to a linear equations system with $N^2$ unknown variables, the real and imaginary parts of the coefficients of the density operator $\rho$ spanned on the canonical base.  Since the number of equations, which is given by the number of measurable probabilities $P^s_j$, is exactly $N^2$ ($N$ bases $\times$ $N$ states) the system can be inverted, provided the determinant of the transformation does not vanish, and consequently, the density operator can be reconstructed.

However, the case $N$ even presents a peculiarity. Casting Eq. (\ref{Psi0}) in the form
\begin{eqnarray}
P^s_j&=&\frac{1}{N}\sum_k\lambda_{k-s}\rho_{k,k}\nonumber\\
&+&\frac{2}{N}\sum_{p<q}\sqrt{\lambda_{p-s}\lambda_{q-s}}\cos(\frac{2\pi}{N}(q-p)j)\Re(\rho_{p,q})
\nonumber\\
&-&\frac{2}{N}\sum_{p<q}\sqrt{\lambda_{p-s}\lambda_{q-s}}\sin(\frac{2\pi}{N}(q-p)j)\Im(\rho_{p,q}),
\nonumber\\
\label{Psi}
\end{eqnarray}
where the coefficients $\rho_{p,q}$ have been separated in its real $\Re(\rho_{p,q})$ and imaginary $\Im(\rho_{p,q})$ parts,
we note that the coefficient $(p-q)$ entering in Eq. (\ref{Psi}) assumes integer values between $N-1$ and $1$, in particular the value $N/2$. Thereby, the argument entering in the sine function in Eq. (\ref{Psi}) becomes an integer multiple of $\pi$ for any value of $j$ and $s$. Thus, the imaginary part of the coefficients $\rho_{p,q}$ with $(p-q)=N/2$ does not appear in the equations system and the previous set of probabilities does not allow us to reconstruct the state completely. Therefore, the previous sets of equidistant states are at most well suited for reconstructing density operators acting onto Hilbert spaces with $N$ odd.  In the following sections we show that this is the case.

\section{Case N=3}
\label{sec4}

Let us now study Eq. (\ref{Psi0}) in the particular case $N=3$. In order to cast this equation in a simpler form we resort to the Fourier transform. Thereby we obtain
\begin{equation}
\tilde P^s_1=\sum_{q=0}^{N-1}\sqrt{\lambda_{q-s}\lambda_{q+1-s}}\rho_{q+1,q},
\end{equation}
where
\begin{equation}
\tilde P^s_1=\sum_{j=0}^{N-1} e^{\frac{2\pi i}{N}j}P^s_j.
\end{equation}
Considering the three possible values of $s$ we generate the following equations system
\begin{equation}
\left(	
\begin{array}{c}
\tilde P^0_1
\\
\tilde P^1_1
\\
\tilde P^2_1
\end{array}
\right)
=
\left(
\begin{array}{ccc}
\sqrt{\lambda_2\lambda_0}  & \sqrt{\lambda_0\lambda_1}  &  \sqrt{\lambda_1\lambda_2} \\
\sqrt{\lambda_1\lambda_2}\ & \sqrt{\lambda_2\lambda_0}  & \sqrt{\lambda_0\lambda_1}  \\
\sqrt{\lambda_0\lambda_1}  &  \sqrt{\lambda_1\lambda_2} & \sqrt{\lambda_2\lambda_0}
\end{array}
\right)
\left(	
\begin{array}{c}
\rho_{1,0}
\\
\rho_{2,1}
\\
\rho_{0,2}
\end{array}
\right).
\label{EQS1N=3}
\end{equation}
This can be separated into two independent sets of equations considering the real and imaginary parts of the coefficients $\rho_{p,q}$ and of $\tilde P^s_1$. These two equations systems are characterized by the same matrix, which in our case turns out to be a circulant matrix, that is each row of the matrix is a cyclic shift of the previous row, being the eigenvalues given by the Fourier transform of one of the rows. Thus, the equations systems can be analytically inverted.

The diagonal coefficients $\rho_{k,k}$ can be obtained by noting that
\begin{equation}
\tilde P^s_0=\sum_{p=0}^2\lambda_{p-s}\rho_{p,p},
\end{equation}
where
\begin{equation}
\tilde P^s_0=\sum_{j=0}^2 P^s_j.
\end{equation}
The three possible values of $s$ lead to the equations system
\begin{equation}
\left(	
\begin{array}{c}
\tilde P^0_0
\\
\tilde P^1_0
\\
\tilde P^2_0
\end{array}
\right)
=
\left(
\begin{array}{ccc}
\lambda_2 & \lambda_0 & \lambda_1 \\
\lambda_1 & \lambda_2 & \lambda_0 \\
\lambda_0 & \lambda_1 & \lambda_2
\end{array}
\right)
\left(	
\begin{array}{c}
\rho_{0,0}
\\
\rho_{1,1}
\\
\rho_{2,2}
\end{array}
\right),
\label{EQS2N=3}
\end{equation}
which turns out to be also defined by a circulant matrix.

Equations systems of Eqs. (\ref{EQS1N=3}) and (\ref{EQS2N=3}) can be considerably simplified by taking $\lambda_1=0$. This particular choice is equivalent to consider the equidistant states in the set ${\cal B}_0(\alpha)$ to be linearly dependent. In this case we have simpler solutions for the non-diagonal coefficients
\begin{equation}
\rho_{1,0}=\frac{\tilde P^0_1}{\sqrt{\lambda_{2}\lambda_0}},~
\rho_{2,1}=\frac{\tilde P^1_1}{\sqrt{\lambda_{2}\lambda_0}},~
\rho_{0,2}=\frac{\tilde P^2_1}{\sqrt{\lambda_{2}\lambda_0}},
\end{equation}
and for the diagonal coefficients we obtain
\begin{eqnarray}
\rho_{0,0}&=&\frac{\tilde P^0_0\lambda_2^2-\tilde P^1_0\lambda_0\lambda_2+\tilde P^2_0\lambda_0^2}{\lambda_0^3+\lambda_2^3},
\nonumber\\
\rho_{1,1}&=&\frac{\tilde P^1_0\lambda_2^2-\tilde P^2_0\lambda_0\lambda_2+\tilde P^0_0\lambda_0^2}{\lambda_0^3+\lambda_2^3},
\nonumber\\
\rho_{2,2}&=&\frac{\tilde P^2_0\lambda_2^2-\tilde P^0_0\lambda_0\lambda_2+\tilde P^1_0\lambda_0^2}{\lambda_0^3+\lambda_2^3}.
\end{eqnarray}

The choice $\lambda_1=0$ is equivalent to make $|\alpha|=|\bar\alpha_\theta|$. It is still possible to select a particular value of $|\bar\alpha_\theta|$  by choosing the value of the phase $\theta$. In the particular case $\theta=\pi$ the inner product $\alpha$ becomes $-1/2$,  $\lambda_0=\lambda_2=3/2$, $(\omega_0)^*=e^{-2\pi i/3}$ and $(\omega_2)^*=e^{2\pi i/3}$. Thereby, the states in the sets ${\cal B}_0(\alpha)$, ${\cal B}_1(\alpha)$ and ${\cal B}_2(\alpha)$  are correspondingly
\begin{eqnarray}
|\alpha_0^0\rangle&=&\frac{1}{\sqrt{2}}(|1\rangle+|0\rangle),\nonumber\\
|\alpha_1^0\rangle&=&\frac{1}{\sqrt{2}}(e^{2\pi i/3}|1\rangle+e^{-2\pi i/3}|0\rangle),\nonumber\\
|\alpha_2^0\rangle&=&\frac{1}{\sqrt{2}}(e^{-2\pi i/3}|1\rangle+e^{2\pi i/3}|0\rangle),\nonumber\\
|\alpha_0^1\rangle&=&\frac{1}{\sqrt{2}}(|2\rangle+|1\rangle),\nonumber\\
|\alpha_1^1\rangle&=&\frac{1}{\sqrt{2}}(e^{2\pi i/3}|2\rangle+e^{-2\pi i/3}|1\rangle),\nonumber\\
|\alpha_2^1\rangle&=&\frac{1}{\sqrt{2}}(e^{-2\pi i/3}|2\rangle+e^{2\pi i/3}|1\rangle),\nonumber\\
|\alpha_0^2\rangle&=&\frac{1}{\sqrt{2}}(|0\rangle+|2\rangle),\nonumber\\
|\alpha_1^2\rangle&=&\frac{1}{\sqrt{2}}(e^{2\pi i/3}|0\rangle+e^{-2\pi i/3}|2\rangle),\nonumber\\
|\alpha_2^2\rangle&=&\frac{1}{\sqrt{2}}(e^{-2\pi i/3}|0\rangle+e^{2\pi i/3}|2\rangle).
\end{eqnarray}
These states are such that the modulo squared of the inner product between any two of them is $1/4$ and consequently they form a SIC-POVM.

\section{Case N odd}
\label{sec4}

The previous results lead us to consider the following transformation of the transition probabilities
\begin{equation}
\tilde P^s_k=\sum_{j=0}^{N-1} e^{\frac{2\pi i}{N}kj}P^s_j
\end{equation}
which when applied onto Eq. (\ref{Psi0}) lead us to the following equation
\begin{equation}
\tilde P^s_k=\sum_{q=0}^{N-1}\sqrt{\lambda_{q+k-s}\lambda_{q-s}}\rho_{q+k,q}.
\label{MASTER}
\end{equation}
For $k$ and $s$ fixed, this equation contains exactly $N$ different coefficients of the density operator. The $N$ values of $s=0,\dots,N-1$ lead to a linear equations system, which for a given $k$ involve the $N$ coefficients of the $k$-th diagonal of $\rho$. In particular, the equations system for $k=0$ involves the coefficients of the main diagonal only. The matrix $M^{(k)}$ defining the equations system has real coefficients only, thus it is possible to generate two equations systems each for the real and imaginary parts of the coefficients of the $k$-th diagonal of $\rho$. Thereby, we obtain one equations system for the $N$ real coefficients of the main diagonal of $\rho$ and $(N-1)/2$ equations systems, each for $N$ complex coefficients belonging to a particular diagonal of $\rho$. This allow us to obtain the value of the $N^2$ real coefficients determining the unknown density matrix $\rho$.

The matrix $M^{(k)}$ is circulant for $k=0,\dots,(N-1)/2$, that is every row of the matrix $M^{(k)}$ is a right cyclic shift of the row above.  This class of matrices can be expressed in the form $M^{(k)}=F^{\dag}D^{(k)}F$, where $F$ is the discrete Fourier transform and $D^{(k)}=diag(\gamma^{(k)}_{0},\gamma^{(k)}_{1},...,\gamma^{(k)}_{N-1})$ is a diagonal matrix which contains the eigenvalues of $M^{(k)}$. The eigenvalues $\gamma^{(k)}_{r}$ can be written as functions of the coefficients $M^{(k)}_{0,t}$ of the matrix $M^{(k)}$, with $t=0,...,N-1$ in the following way
\begin{equation}
\gamma^{(k)}_{r}=\sum_{m=0}^{N-1}M^{(k)}_{0,m}e^{-\frac{2\pi i}{N}rm},~~~r=0,\dots,N-1.
\end{equation}

The inverse of the matrix $M^{(k)}$, provided $M^{(k)}$ is non-singular, is also circulant and is given by $[M^{(k)}]^{-1}=F^{\dag}[D^{(k)}]^{-1}F$, where in this case the inverse of the $D^{(k)}$ matrix is given by
\begin{equation}
[D^{(k)}]^{-1}=diag(\frac{(\gamma^{(k)}_{0})^*}{|\gamma^{(k)}_{0}|^2},\frac{(\gamma^{(k)}_{1})^*}{|\gamma^{(k)}_{1}|^2},\dots,\frac{(\gamma^{(k)}_{N-1})^*}{|\gamma^{(k)}_{N-1}|^2}).
\end{equation}
Therefore, the elements of the first row that defines $[M^{(k)}]^{-1}$ are given by
\begin{equation}
[M^{(k)}]^{-1}_{0,l}=\frac{1}{N}\sum_{r=0}^{N-1}\frac{(\gamma^{(k)}_{r})^*}{|\gamma^{(k)}_{r}|^2}e^{\frac{2\pi i}{N}lr}, l=0,\dots,N-1.
\end{equation}
Considering Eq. (\ref{MASTER}) we obtain that the coefficients of the density matrix are related to the probabilities $P_i^s$ through the relation
\begin{equation}
\rho_{k+q,q}=\frac{1}{N}\sum_{r,l,j=0}^{N-1} \frac{(\gamma^{(k)}_{r})^*}{|\gamma^{(k)}_{r}|^2} e^{\frac{2\pi i}{N}[(l-q)r+kj]}P^l_j,
\end{equation}
where the eigenvalues $\gamma^{(k)}_{r}$ are given by
\begin{equation}
\gamma^{(k)}_{r}=\sum_{m=0}^{N-1}\sqrt{\lambda_{m+k}\lambda_{m}}e^{-\frac{2\pi i}{N}mr}.
\end{equation}
Let us consider for instance the case $\theta=0$. In this case we obtain that the values of $\lambda_k$ are given by
\begin{eqnarray}
\lambda_0&=&1+(N-1)|\alpha|
\nonumber\\
\lambda_k&=&1-|\alpha|~~~~~~~k=1,\dots,N-1
\end{eqnarray}
and thus the eigenvalues $\gamma^{(k)}_{r}$ vanish only for $|\alpha|=0$, that is when the equidistant states in ${\cal B}_0(\alpha)$ are mutually orthogonal. A similar result holds in the case $\theta=\pi$.

\section{Conclusions}
\label{sec6}

In the previous sections we studied the possibility of applying equidistant states to the problem of determining unknown states of finite dimensional quantum systems. We have shown that when the dimension $N$ of the quantum system is odd, it is possible to formulate a scheme for quantum tomography. This scheme has two attributes that place it between standard quantum tomography and POVM-based quantum tomography. Firstly, the new scheme requires the measurement of  $N^2$ transition probabilities, as few as in the case of MUB or SIC quantum tomography. Secondly, the new scheme requires the solution of a system of linear equations. Thus, it requires to perform an inversion as in the case of standard quantum tomography. However, this inversion reduces to the problem of inverting $(N+1)/2$ matrices each one of $N\times N$. Since these matrices are circulant analytical expressions are known for their eigenvalues and eigenvectors.

An interesting feature of the equidistant states based quantum tomography is that each one of the $(N+1)/2$ matrices provides information about a particular diagonal of the density matrix to be determined. Thereby, errors in the reconstruction are constrained within diagonals and do not propagate to other diagonals. Finally, let us note that the value of $\alpha$ can be freely chosen as long as the eigenvalues $\gamma^{(k)}_{r}$ do not vanish.

\acknowledgments

This work was supported by Grants ICM P06-067-F, CONICyT FB0824 and FONDECyT N$^{\text{\underline{o}}}$ 1080383.

\end{document}